\documentclass{kluwer}
\newcommand{\Msolar}{\mbox{\,$\rm M_{\odot}$}}        
\begin{document}
\begin{article}
\newcommand{\Msun}{\mbox{\,$\rm M_{\odot}$}}        
\newcommand{\Rsun}{\mbox{\,$\rm R_{\odot}$}}        
\newcommand{\dg}{$^{\circ}$}
\newcommand{\kms}{\mbox{\,km\,s$^{-1}$}}            
\newcommand\Teff{$ {\rm T_{eff}}$}
\newcommand\teff{$ {\rm T_{eff}}$}
\newcommand\logM{$\log {\rm M}$}
\newcommand\logt{$\log {\rm T_{eff}}$}
\newcommand\logg{$\log {\rm g}$}
\newcommand\loghe{${\rm \log{\frac{n_{He}}{n_{H}}}}$}
\begin{opening}
\title{Atmospheric parameters and abundances of sdB stars}            

\author{U. \surname{Heber}\email{heber@sternwarte.uni-erlangen.de}}
\author{H. \surname{Edelmann}} 
\institute{Dr. Remeis-Sternwarte, Universit\"at
Erlangen-N\"urnberg, Bamberg, Germany}                               




\runningtitle{Atmospheric parameters and abundances of sdB stars}
\runningauthor{Heber et al.}



\begin{abstract}
We summarize recent results of quantitative spectral analyses using NLTE and
metal line-blanketed LTE model atmospheres. 
Temperatures and gravities derived for hundreds of sdB stars are now 
available and allow us to investigate systematic uncertainties of \teff, 
log~g scales and to test the theory of stellar evolution and pulsations.
Surface abundance patterns of about two dozen sdB stars are surprisingly 
homogenous. In particular the iron abundance is almost solar for most sdBs. We
highlight one iron deficient and three super metal-rich sdBs, a challenge to 
diffusion theory. SdB stars 
are slowly rotating stars unless they are in close binary systems which is
hard to understand if the sdB stars were formed in merger events. The only
exception is the pulsator PG~1605+072 rotating at v$\sin\,i=39km/s$. 
Signatures 
of stellar winds from sdB stars have possibly been found. 

\end{abstract}




\end{opening}

The theories for the
evolution and pulsations of sdB stars, for diffusion processes 
in their envelopes, and for winds from their surfaces have made enormous 
progress during recent years. However, they need to be tested 
observationally and we need to know
the sdB atmospheric parameters (\teff, log~g), surface abundances, rotation
velocities, and mass loss rates 
accurately. Quantitative spectral analyses using LTE and 
NLTE model atmospheres are required to achieve this goal.

The first quantitative analyses of sdB spectra 
(Sargent \& Searle, 1966) already revealed that helium is strongly deficient 
(by a factor of 100) due to gravitational settling (Greenstein et al., 1967) 
Metal abundances were also found be peculiar (Baschek et al., 1972).
In their pioneering papers, Newell (1973) and 
Greenstein \& Sargent (1974) presented temperatures and gravities of
large numbers of sdB stars and related objects and concluded that the sdB 
stars form an 
extension of the horizontal branch, with \teff$^4$/g=constant.
UBV colours were used as \teff\ and 
Balmer line widths as log~g indicators. With the advent of the IUE satellite
the spectral energy distribution was used 
to determine \teff\ while gravity was
derived from Balmer line profiles. High-resolution IUE UV spectra 
allowed the abundances of many metals to be measured 
(see Heber 1998 for a review). New generations of model atmospheres and high 
quality spectra have become available in the mean time which allow accurate
quantitative analyses even of quite faint sdB stars.  

\subsection*{Model atmospheres and atmospheric parameters}

\begin{figure}
\vspace{10.5cm}
\includegraphics{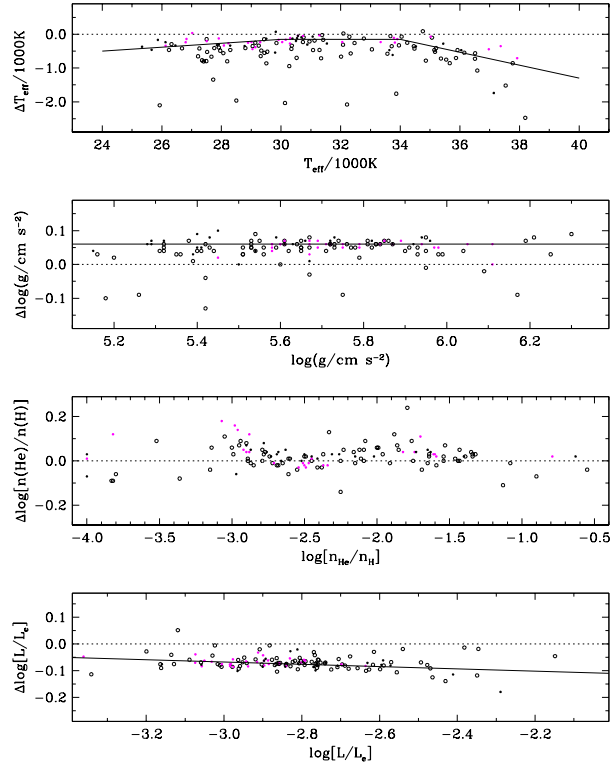}
	\caption[]{Comparison of atmospheric parameters derived from LTE and
NLTE model atmosphere analyses ($\Delta$=LTE-NLTE).
	}\label{diff}
\end{figure}

\begin{figure}
\vspace{7.0cm}
\includegraphics{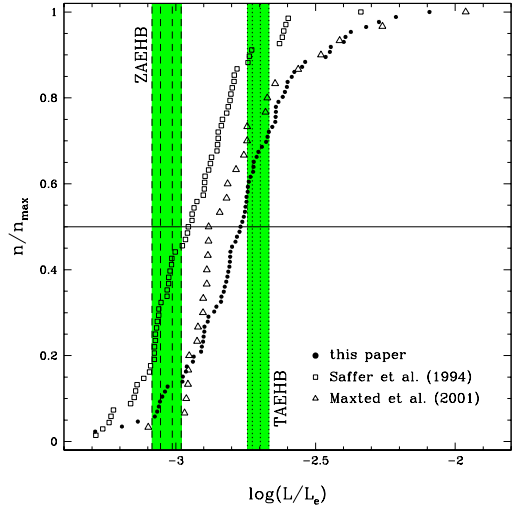}
	\caption[]{Plot of (normalized) numbers of stars versus luminosity (with respect to the Eddington
   luminosity $L_e$).
  Additionally are plotted the results of Saffer et al. (1994, open squares) 
  and Maxted et al. (2001, open triangles).
   The  dashed vertical lines denote the ZAEHB, and the dotted vertical lines
   the TAEHB for metallicities of [M/H]=0.00, $-$0.47, $-$1.48, and $-$2.22, respectively 
   from the left to the right (Dorman, Rood \& O'Connell, 1993).
	}\label{logl_le_n}
\end{figure}

Saffer et al. (1994) introduced 
the method of Balmer and helium line profile matching which has now become the 
Standard analysis method in the field. Synthetic spectra calculated
from LTE and NLTE model atmospheres, respectively, are matched to the 
observations by a 
$\chi^2$ method and \teff, g, and He abundance are fitted simultaneously.
Different model atmospheres have been used. Saffer et al. used metal-free 
LTE models, whereas Heber et al. (1999, 2000), Maxted et al. (2001), and 
Edelmann et al. (2003) used metal line blanketed LTE and metal-free NLTE 
model atmospheres to calculate synthetic spectra.
It has been realized (e.g. Saffer et al. 1994) that results from different
methods can be quite different. Also the use of different atmospheric models
can result in systematic differences (Heber et al., 2000). For the sample of
Edelmann et al. (2003) we compared the results from two different model 
atmosphere sets, i.e. metal-line blanketed LTE model atmospheres versus 
metal-free NLTE atmospheres (see Fig. \ref{diff}). Effective temperatures 
derived from NLTE models are higher by only a few hundred Kelvin up to
\teff=35000\,K. NLTE gravities are lower by about 0.06~dex, which translates
into a luminosity offset of about the same amount increasing somewhat with
increasing luminosity.  
A comparison of results for individual stars 
is as yet possible only for a few stars common to the samples. 
We can, however, compare the samples in a global sense using their cumulative 
luminosity functions.
In Fig. \ref{logl_le_n} we plot these functions for our sample and those of
Saffer et al. (1994) and Maxted et al. (2001).
The luminosity is expressed in units of the Eddington luminosity. 
Additionally the positions of the ZAEHB and the TAEHB are indicated. 
Since the metallicity of the stars is unknown, we have plotted models for 
various metallicities.
As can be seen from Fig. \ref{logl_le_n}, the overall shape of the
cumulative luminosity function is similar for all three samples.
However, there is an offset of about 0.2 dex in luminosity between the  
Saffer et al. (1994) sample and Edelmann et al. (2003), 
whereas the relation for the Maxted et al. and Edelmann et al. samples are in 
better agreement.
Possible reasons for the discrepancy are
different observations, the different
synthetic spectra used in the analysis, or both.

\subsection*{Helium abundances and the $^3$He isotopic anomaly}

\begin{figure}
\vspace{6.5cm}
\includegraphics{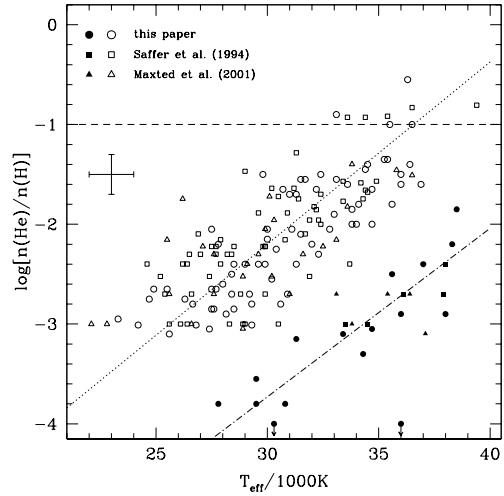}
	\caption[]{Plot of the helium abundance versus effective 
	tem\-pe\-ra\-tu\-re. 
  Additionally the results of Saffer et al. (1994, squares) 
  and Maxted et al. (2001, triangles) are plotted.
  The dotted line indicates the regression line for the 
  bulk 
  of the sdB stars (open symbols)
  and the dashed-dotted line shows the regression line
  for the peculiar sdB stars
  (filled symbols). 
  The dashed horizontal line denotes the solar helium abundance. 
}
 	\label{nhe_teff}
\end{figure}

A correlation between the
photospheric helium content and \teff\ (Fig.~\ref{nhe_teff}) and log~g  
becomes apparent. 
The larger the
effective temperature and /or gravity, the larger is the helium abundance.
However, for sdB stars, $T_{\rm eff}$ and gravity are strongly connected 
and a plot of helium abundance versus luminosity does not reveal 
any correlations.
Additionally, a population of stars with very low helium abundances was 
identified.
These stars clearly separate from the bulk (see Fig.~\ref{nhe_teff}) providing
evidence that surface abundances of sdB stars 
are not a simple function of their position in the HR diagram. It rules out 
time-independent diffusion models and points to a dependence on the star's 
history.


%

Because gravitational settling depends on atomic mass
we expect the $^3$He/$^4$He ratio to increase slowly with time. Therefore the 
isotopic ratio can potentially be calibrated by diffusion models as an age
indicator.

It can be measured best from the isotopic 
line shift of the He~I 6678\AA\ line (0.5\AA). 
He~I 5876\AA\ provides an important check because its 
line shift is almost negligible (0.04\AA).
Because of its very low abundance, the $^3$He isotope can not be detected 
in normal B stars.  The presence of $^3$He
in the spectrum of an sdB star was first discovered in the case of SB~290
(Heber, 1991), for which the measured line shift is larger than for any
other known $^3$He star, indicating that $^4$He is almost entirely replaced 
by $^3$He in SB~290. $^3$He was also discovered in Feige~36 (Edelmann et al., 
1999). We have carried out a search for the $^3$He anomaly 
amongst the brightest sdB stars and discovered two more $^3$He sdBs:
BD+48\dg2721 and PG~0133+114 and found the $^3$He anomaly to be rare, i.e. it
occurs in less than 5\% of our programme stars. 

\begin{figure}
\vspace{6.9cm}
\includegraphics{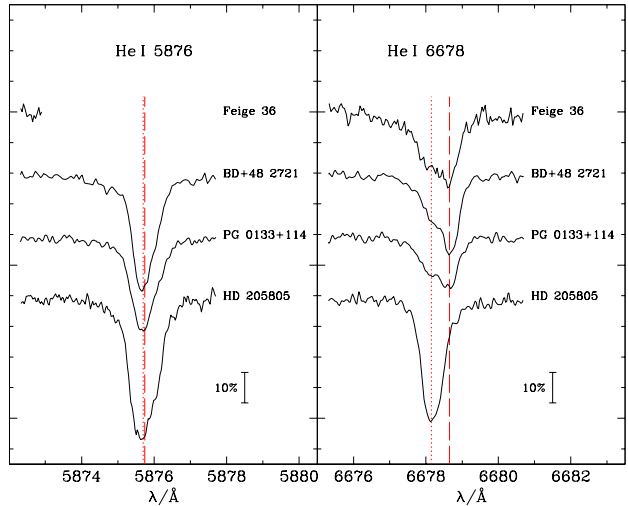}
	\caption[]{Observed line profiles of He~I 5876\AA\ and 6678\AA\
	in the $^3$He sdB stars Feige~36, BD+48\dg2721, and PG~0133+114 and
        the normal sdB HD~205805. The laboratory line positions for  $^3$He
        (dotted) and  $^4$He (dashed) are indicated. Note that the $^3$He
        line is stronger than the $^4$He component.}
\end{figure}

\subsection*{metal abundances}

\begin{figure}
\vspace{9.0cm}
\includegraphics{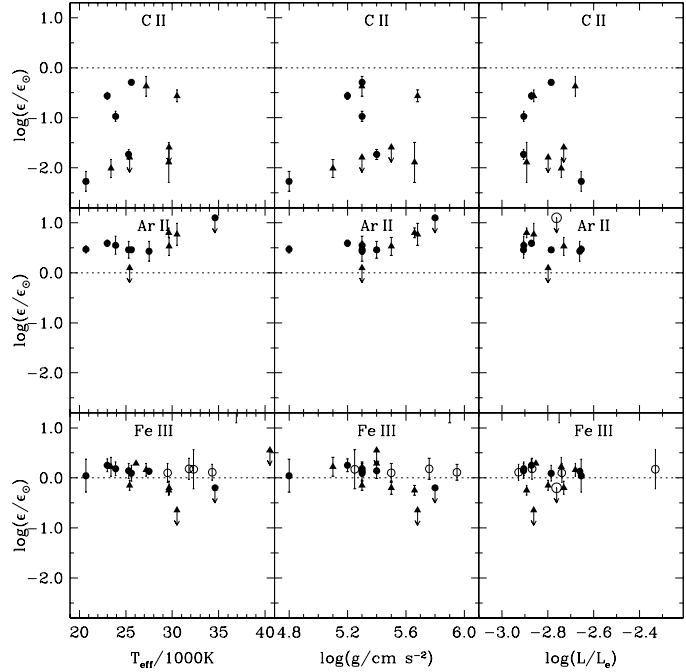}
	\caption[]{LTE abundances for C\,II, Ar\,II and Fe\,III. Filled
circles mark stars that are not known to be pulsating or radial velocity
variable. Pulsating
stars are plotted as open circles, radial velocity variables as filled
triangles. Upper limits are marked by arrows pointing downwards. 
The dashed line denotes the solar abundance.}\label{carfe}
\end{figure}

We have determined abundances of C, N, O, Mg, Al, Si, S, Ar, and Fe 
of pulsating (Heber et al. 1999, 2000) 
and non-pulsating stars (Napiwotzki et al. 2001; Edelmann et al., in prep.) 
covering a wide range of atmospheric parameters, i.e. 
\teff~=~20\,000~K~\ldots~40\,000~K and log~g~=~4.8~\ldots~6 in order to
search for possible trends for the elemental abundances with atmospheric
parameters.  Fig. \ref{carfe}
displays the results for C\,II, Ar\,II and Fe~III as an illustration.
While carbon and oxygen display a large scatter, the abundances of other 
elements (most notably N) are quite homogenous and no trends with the 
\teff, log~g or L/L$_{e}$ become apparent (see Fig.~\ref{carfe}). 
While many elements have subsolar abundances (by 0.5 to 1.0 dex), iron 
is nearly solar in most cases. There are, however, three stars, for which no 
iron lines could be detected and upper limits for Fe were determined only.
HE~1047-0436 is deficient in iron by 0.6 dex or more (Napiwotzki et al., 2001).
This star has a white dwarf companion. Except for its low iron content 
the abundance pattern is typical for sdB stars. 
Comparing radial velocity variable stars and
non-variable ones we do not find any significant differences except for 
HE~1047-0436. 
The same holds for the comparison of pulsating and
non-pulsating stars (see Fig. \ref{carfe}).

\begin{figure}
\vspace{6.0cm}
\includegraphics{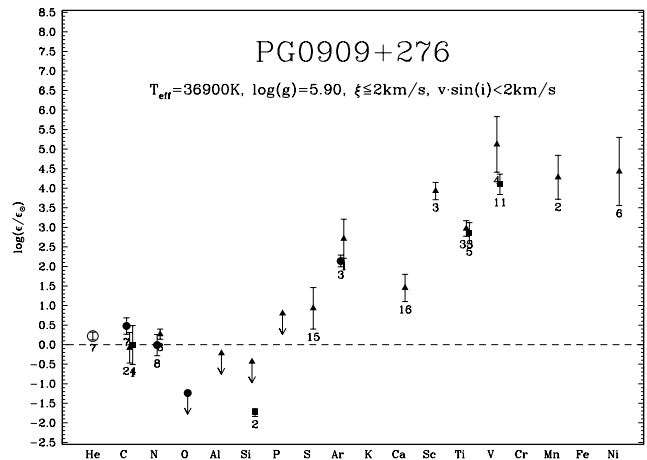}
	\caption[]{LTE abundances for PG~0909+276 relative to solar
composition (dashed line).Abundances derived from neutral elements are shown
as open circles, from doubly ionized ones as filled triangles, and from
triply ionized ones as filled squares. The numbers at each error bar indicated 
the number of lines used.}\label{super}
\end{figure}

Three stars in our sample (not included in Fig.~\ref{carfe}) have very
peculiar abundances, one example is shown in Fig.~\ref{super}. Their optical
spectra show lots of lines mostly from Ca III, Ti III/IV and V III/IV
but no iron lines. These spectral lines have not be seen in any other sdB star before.
These elements are strongly enriched with respect to the sun (1000 times for
V and 10\,000 times for Ti).

%

\subsection*{Rotation and stellar winds from sdB stars}

Projected rotation velocities derived from optical spectra 
provide important information about the angular momentum
evolution of the stars. A large fraction of sdB stars may have been formed by 
mergers of two He white dwarfs (Han et al., 2003) during which angular
momentum is transferred. Hence we expect such sdB stars to be born with a high
surface rotation.     
However, the spectra of most sdB stars are very sharp lined indicating
that they are slow rotators (v $\sin\,i<5km/s$). However, rapidly 
rotating stars  
such as HS0705+6700 (Drechsel et al. 2001) are found amongst close binary
systems, their rotation being caused by tidal locking to the binary orbit.
The enigmatic pulsator PG~1605+072 is the only sdB star known today that is 
rotating (v$\sin\,i=39km/s$) but not in a close binary system. Is it a merger?
The masses of sdB stars formed by mergers tend to be higher than the
canonical mass (Han et al. 2003). 
Hence an independent measurement of its mass would be tale
telling. This can possibly be achieved by asteroseismology (see O'Toole et al.
this meeting). 

Stellar winds have
frequently been suggested as an explanation for the helium abundances.
The first realistic calculations 
have been carried out by Fontaine \& Chayer (1997) and Unglaub \& Bues 
(2001).
The observed He abundances can indeed be explained 
if a mass loss rate of 10$^{-12}$ to 10$^{-14}$\Msolar/yr is adopted. 
Yet it is not clear whether diffusion models incorporating mass 
loss can explain the metal abundance anomalies, at the same time as the He 
abundance.

Mass loss has been detected in some ``low gravity'' sdO stars but not in 
any sdB star.
Since sdBs are less luminous we expect the mass loss rates to be lower 
than in sdO stars and therefore harder to detect. Indeed, up to now there 
is no observational proof for mass loss and, therefore, the mass loss rate 
is still a free parameter in diffusion models. 

First hints about stellar winds came from the 
quantitative analysis of  H$\alpha$ line profiles of 40 sdB stars 
(Heber et al. 2003)
A comparison of synthetic NLTE H$\alpha$ line 
profiles to the observations revealed perfect matches for all stars 
except for the
four most luminous sdBs. Heber et al. (2003) speculate that the 
peculiar H$\alpha$ lines are a
signature of a stellar wind. Because the mass loss rate is expected to
increase with increasing luminosity, the most luminous sdBs are most
likely to show wind signatures. Sophisticated wind models such as presented 
by Vink at this meeting and UV observations are required to prove or
disprove this conjecture and to determine mass-loss rates.  


\subsection*{Conclusion}

The field of research into EHB stars is rapidly developing. 
Quantitative spectral analyses are progressing very well, both in quantity 
(number of stars) as in quality (fine abundance analyses). New subjects
have been addressed recently, such as stellar winds from sdB stars.
The next step has already been undertaken: a homogenous set of high quality,
high resolution spectra for a large sample of sdB stars drawn from 
the ESO SPY survey are analysed (preliminary results are presented by 
Lisker et al. at this conference). A thorough investigation of the 
temperature and gravity scale, however, has still to be done.
Finally, the role of magnetic fields in sdB stars is unexplored up
to now but can now be tackled with modern telescopes and instruments (e.g.
ESO VLT + FORS).



\end{article}
\end{document}